\documentclass[sigconf]{acmart}

\copyrightyear{2021} 
\acmYear{2021} 
\setcopyright{acmcopyright}
\acmConference[WSDM '21] {Proceedings of the Fourteenth ACM International Conference on Web Search and Data Mining}{March 8--12, 2021}{Virtual Event, Israel}
\acmBooktitle{Proceedings of the Fourteenth ACM International Conference on Web Search and Data Mining (WSDM '21), March 8--12, 2021, Virtual Event, Israel}
\acmPrice{15.00}
\acmDOI{10.1145/3437963.3441835}
\acmISBN{978-1-4503-8297-7/21/03}

\usepackage{booktabs} 
\usepackage{multirow}
\usepackage{graphicx} 
\usepackage{subfig} 
\usepackage{threeparttable} 
\usepackage{amsmath} 
\usepackage{amsfonts}
\usepackage{url}

\begin{document}
\fancyhead{}
\title{Heterogeneous Hypergraph Embedding for  Graph Classification }

\author{Xiangguo Sun}
\affiliation{%
  \institution{Southeast University}
  \country{China}
}
\email{sunxiangguo@seu.edu.cn}

\author{Hongzhi Yin}
\authornote{Hongzhi Yin is the second corresponding author}
\affiliation{%
  \institution{The University of Queensland}
  \country{Australia}
}
\email{h.yin1@uq.edu.au}

\author{Bo Liu}
\authornote{Bo Liu is the first corresponding author}
\affiliation{%
  \institution{Southeast University}
  \country{China}
}
\email{bliu@seu.edu.cn}

\author{Hongxu Chen}
\affiliation{%
  \institution{University of Technology Sydney}
  \country{Australia}
  }
\email{hongxu.chen@uts.edu.au}

\author{Jiuxin Cao}
\affiliation{%
  \institution{Southeast University}
  \country{China}
  }
\email{jx.cao@seu.edu.cn}

\author{Yingxia Shao}
\affiliation{%
  \institution{Beijing University of Posts and Telecommunications}
  \country{China}
  }
\email{shaoyx@bupt.edu.cn}

\author{Nguyen Quoc Viet Hung}
\affiliation{%
  \institution{Griffith University}
  \country{Australia}
  }
\email{quocviethung1@gmail.com}

\renewcommand{\shortauthors}{Sun and Yin, et al.}

\begin{abstract}
  \par Recently, graph neural networks have been widely used for network embedding because of their prominent performance in pairwise relationship learning. In the real world, a more natural and common situation is the coexistence of pairwise relationships and complex non-pairwise relationships, which is, however, rarely studied. In light of this, we propose a graph neural network-based representation learning framework for heterogeneous hypergraphs, an extension of conventional graphs, which can well characterize multiple non-pairwise relations. Our framework first projects the heterogeneous hypergraph into a series of snapshots and then we take the Wavelet basis to perform localized hypergraph convolution. Since the Wavelet basis is usually much sparser than the Fourier basis, we develop an efficient polynomial approximation to the basis to replace the time-consuming Laplacian decomposition. Extensive evaluations have been conducted and the experimental results show the superiority of our method. In addition to the standard tasks of network embedding evaluation such as node classification, we also apply our method to the task of spammers detection and the superior performance of our framework shows that relationships beyond pairwise are also advantageous in the spammer detection. 
  
  To make our experiment repeatable, source codes and related datasets are available at \url{https://xiangguosun.mystrikingly.com}. 
\end{abstract}

\ccsdesc[500]{Computing methodologies~Neural networks}

\keywords{heterogeneous hypergraph; wavelet neural networks; graph neural networks; spammer detection}

\maketitle



\begin{figure}[t]
\centering
\includegraphics[width=0.4\textwidth]{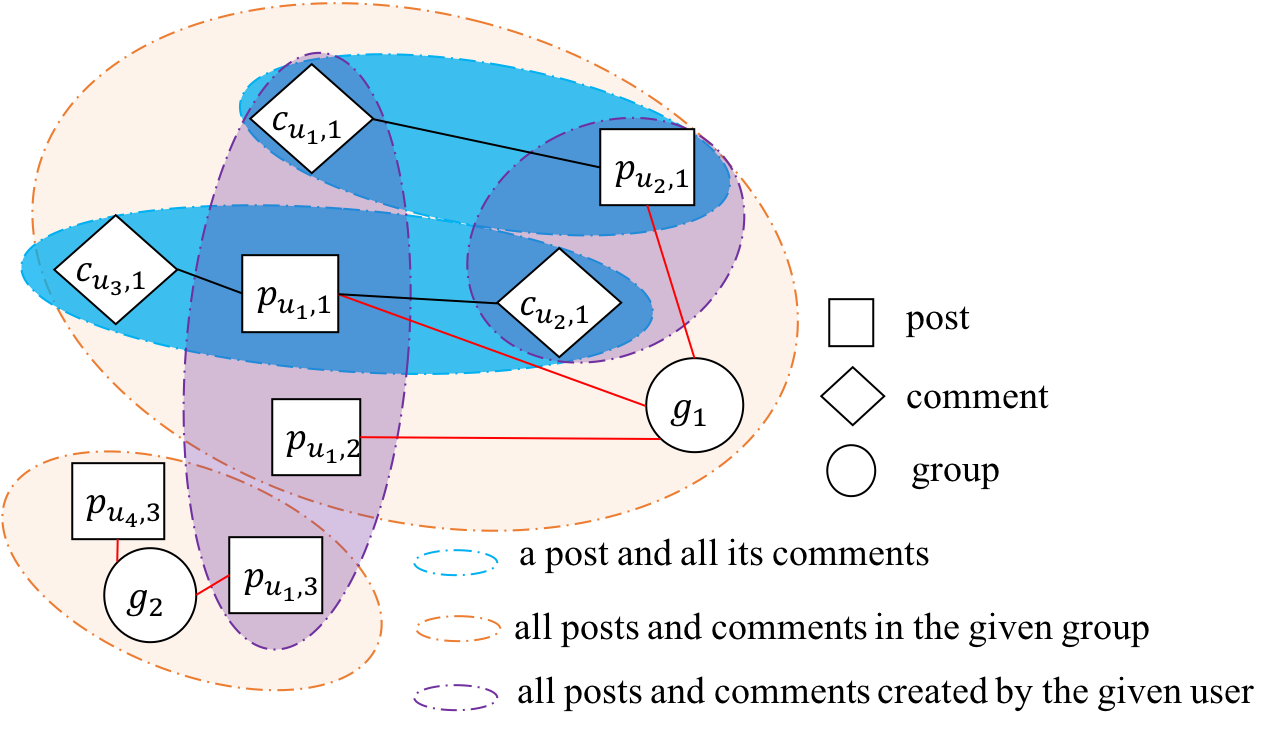}
\caption{A heterogeneous hypergraph on online social forums. 
There are several types of hyperedges, including all posts and comments created by a specific user (the purple circles), all posts and comments in the same group (the orange circles), and a post with all its comments (the blue circles). 
}
\label{fig:hypergraph}
\end{figure}
\section{Introduction}\label{sec:intro}
Recently, Graph Neural Networks (GNNs) have attracted significant attention because of their prominent performance in various machine learning applications \cite{yin2019social, yin2017spatial, chen2020social}. Most of these methods focus on the pairwise relationships between objects in the constructed graphs.In many real-world scenarios, however, relationships among objects are not dyadic (pairwise) but rather triadic, tetradic, or higher. Squeezing the high-order relations into pairwise ones leads to information loss and impedes expressiveness. 

To overcome this limitation, hypergraph \cite{zhou2007learning} has been recently proposed and achieved remarkable improvments\cite{Bretto2013}. Hypergraphs allow one hyperedge to connect multiple nodes simultaneously so that interactions beyond pairwise relations among nodes can be easily represented and modelled. Figure \ref{fig:hypergraph} is an example of heterogeneous hypergraph reflected in online social forums. Specifically, $g_1, g_2$ are two different social groups, $p_{u,i}$ denotes the i-th post created by user $u$, and $c_{u,i}$ is the i-th comment created by user $u$. There exist both pairwise relationships and more complex relationships.

\begin{figure}[t]
\centering
\includegraphics[width=0.3\textwidth]{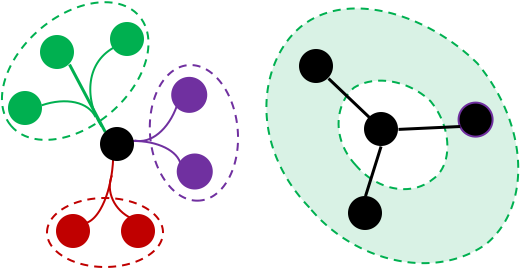}
\caption{Comparison between the hypergraph (left) and the simple graph (right) which share the same type of nodes. A node in the hypergraph usually has different types of neighbors even when they are the same type of nodes, while a node in the homogeneous simple graph usually has one type of neighbors. 
}
\label{fig:nei}
\end{figure}

Despite the potentials of hypergraphs, only a few works shift attentions to representation learning on hypergraphs. Earlier works, \cite{zhang2018dynamic,zhao2018personality,zheng2018novel} mostly design a regularizer to integrate hypergraphs into specific applications, which are domain-oriented and hard to be generalized to other domains. Recently, some researches \cite{feng2019hypergraph, jiang2019dynamic} try to design more universal learning models on hypergraphs. For example, Yadati et al. \cite{yadati2019hypergcn} transform a hypergraph into simple graphs and then use convolutional neural networks for simple graphs to learn node embeddings. Tu et al. \cite{tu2018structural} learn the embeddings of a hypergraph to preserve the first-order and the second-order proximities. Zhang et al. \cite{DBLP:conf/iclr/ZhangZ020} take the analogy with natural language processing and learn node embeddings by predicting hyperedges. However, they mostly focus on the same type of entities, or apply the concept of heterogeneous simple graphs directly to hypergraphs. But there are key differences between heterogeneous simple graphs and heterogeneous hypergraphs. Even for those homogeneous simple graphs like Figure \ref{fig:nei}, the same type nodes may also be connected according to different semantics that are represented by different types of hyperedges, making the hypergraph heterogeneous (\textbf{\textit{challenge 1}}). 

Recently, graph neural networks have show great power on graph learning, traditional GNN based methods take the assumption that information should be aggregated via point-to-point channel iteratively because links in simple graphs are pairwise. As shown in Figure \ref{fig:nei}, messages can be directly aggregated from one-hop neighbors in the simple graph. However, message diffusion is more complex on hypergraphs. It should be first aggregated within the same hyperedge, and then aggregated over all hyperedges connecting to the target node. This difference makes traditional GNN-based methods unfit for hypergraphs  (\textbf{\textit{challenge 2}}). 
 
\par To address \textbf{\textit{challenge 1}}, we first extract simple graph snapshots with different meta-path, then we construct several hypergraph snapshots on these simple graphs according to hyperedge types. After the decomposition, each snapshot is homogeneous, and they can also be easily calculated in parallel, making the model scalable to large datasets. To address \textbf{\textit{challenge 2}}, we design a hypergraph convolution by replacing the Fourier basis with a wavelet basis. Compared with methods in the vertex domain, this spectral method does not need to consider the complex message passing pattern in hypergraphs and can also perform localized convolution. Since the Wavelet basis is much sparser than the Fourier basis, it can be efficiently approximated by polynomials without Laplacian decomposition. In summary, the main contributions of this paper are as follows:
\begin{itemize}
	\item We focus on the heterogeneity of hypergraphs, and address the problem via simple graph snapshots and hypergraph snapshots according to different meta-paths and hyperedge types respectively. 
	\item We propose a novel heterogeneous hypergraph neural network to perform representation learning on heterogeneous hypergraphs. To avoid the time-consuming Laplacian decomposition, we introduce a polynomial approximation-based Wavelet basis to replace the traditional Fourier basis. To the best of our knowledge, we are the first paper to introduce wavelets in hypergraph learning. 
	\item Extensive evaluations have been conducted, and the experimental results on three datasets demonstrate the significant improvement of our model over six state-of-the-art methods. Even in a sparsely labeled situation, our method still keeps ahead. We also evaluate the performance of our model in the task of spammer detection, and it produces much higher accuracy than three competitive baselines, which further demonstrates the superiority of hypergraph learning. 
\end{itemize}


\section{Preliminary and Problem Formulation}\label{sec:pre}
In this section, we first introduce some necessary definitions and notations, and then formulate the problem of heterogenous hypergraph embedding.
\newtheorem{definition3}{Definition}



\begin{figure}[t]
\centering
\subfloat[Simple graph snapshots]{
\label{fig:sim_snap}
\includegraphics[width=0.25\textwidth]{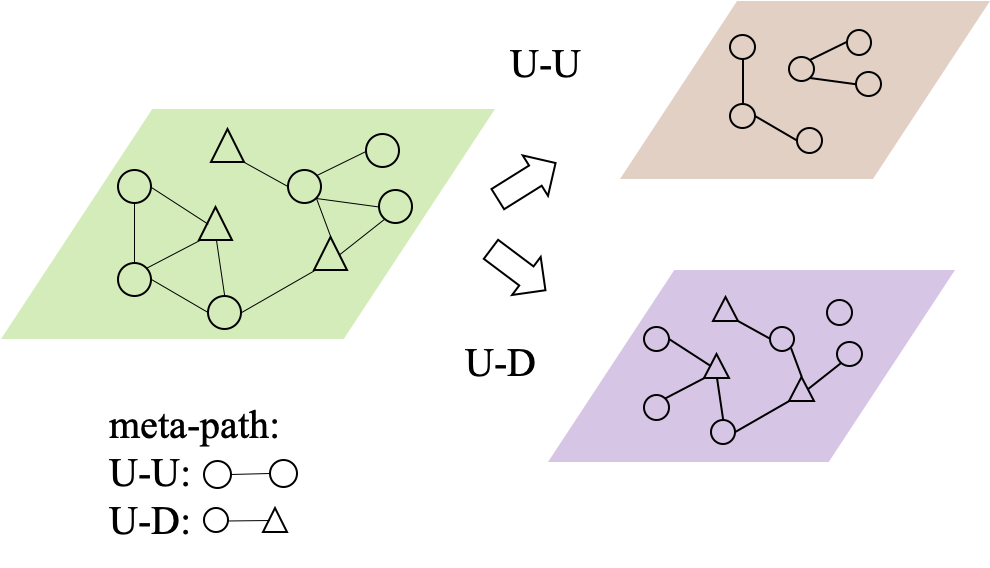}%
}
\subfloat[Hypergraph snapshots]{
\label{fig:hyper_snap}
\includegraphics[width=0.23\textwidth]{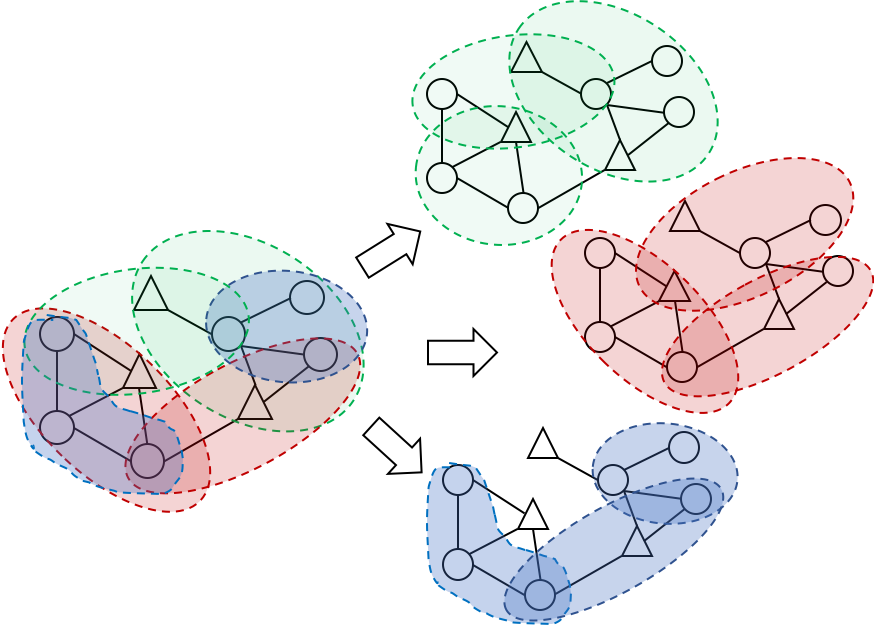}%
}
\caption{Snapshots generation for heterogeneous simple graphs and hypergraphs}
\label{fig:snap}
\end{figure}

\begin{definition3}
\textbf{(Simple Graph Snapshots)}.
According to the selected meta-paths, we can extract the corresponding subgraphs from the original heterogeneous simple graph. Take Figure \ref{fig:sim_snap} as an example, we represent the social network with users (U), and departments (D) as nodes, where edges represent friendships (U-U), and affiliation (U-D) relationships. We extract paths according to meta-path U-U and meta-path U-D, then we can generate two subgraphs as two snapshots for the simple graph.
\end{definition3}

\begin{definition3}
\textbf{(Heterogeneous Hypergraph)}.
A heterogeneous hypergraph can be defined as $\mathcal{G}=\{\mathcal{V},\mathcal{E},\mathcal{T}_v,\mathcal{T}_e,\mathbf{W}\}$. Here $\mathcal{V}$ is a set of vertices, and $\mathcal{T}_v$ is the vertex type set. $\mathcal{E}$ is a set of hyperedges, and $\mathcal{T}_e$ is the set of hyperedge types.  When $|\mathcal{T}_v|+|\mathcal{T}_e|> 2$, the hypergraph is heterogeneous. Hypergraphs allow more than two nodes to be connected by a hyperedge. For any hyperedge $e \in \mathcal{E}$, it can be denoted as $\{v_i,v_j,\cdots,v_k\} \subseteq \mathcal{V}$. We use a positive diagonal matrix $\mathbf{W} \in \mathbb{R}^{|\mathcal{E}| \times |\mathcal{E}|}$ to denote the hyperedge weights. The relationship between nodes and hyperedges can be represented by an incidence matrix $\mathbf{H} \in \mathbb{R}^{|\mathcal{V}|\times |\mathcal{E}|}$ with entries defined as:
\begin{equation*}
\begin{split}
\mathbf{H}(v,e)=\left\{
             \begin{array}{ll}
             1,& \text{if } v \in e \\
             0, &\text{otherwise} 
             \end{array}
\right.
\end{split}
\end{equation*}
Let $\mathbf{D}_v \in \mathbb{R}^{|\mathcal{V}|\times |\mathcal{V}|}$ and $\mathbf{D}_e \in \mathbb{R}^{|\mathcal{E}|\times |\mathcal{E}|}$ denote the diagonal matrices containing the vertex and hyperedge degrees respectively, where $\mathbf{D}_v(i,i)=\sum_{e \in \mathcal{E}}\mathbf{W}(e)\mathbf{H}(i,e)$ and $\mathbf{D}_e(i,i)=\sum_{v\in \mathcal{V}}\mathbf{H}(v,i)$. Let $\mathbf{\Theta}=\mathbf{D}_v^{-\frac{1}{2}}\mathbf{H}\mathbf{W}\mathbf{D}_e^{-1}\mathbf{H}^T\mathbf{D}_v^{-\frac{1}{2}}$, then the hypergraph Laplacian is $\Delta=\mathbf{I}-\mathbf{\Theta}$.
\end{definition3}

\begin{figure*}[h]
\centering
\includegraphics[width=0.9\textwidth]{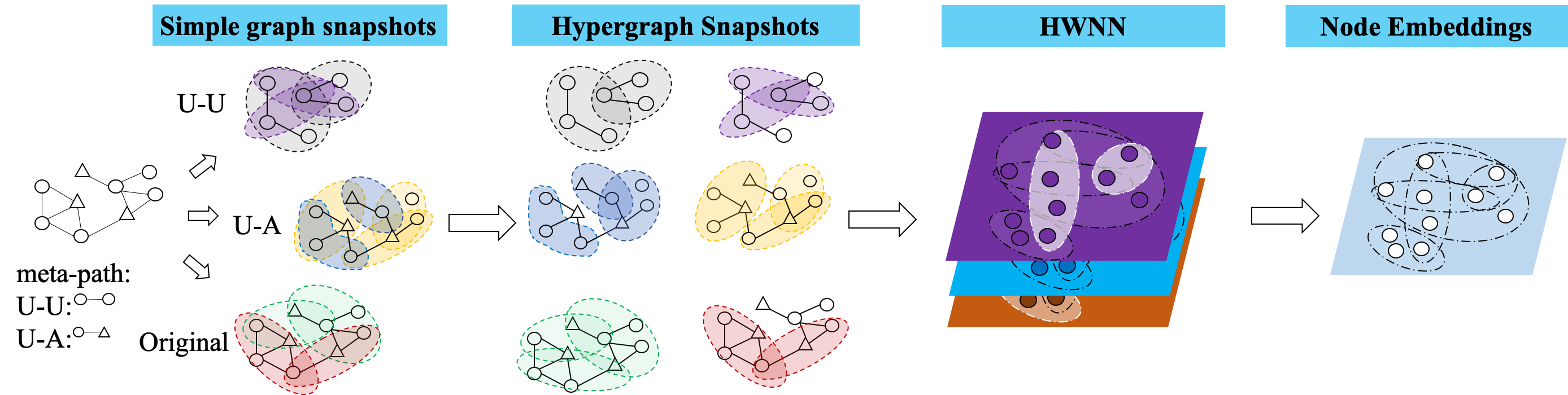}
\caption{The flowchart of our framework.}
\label{fig:framework}
\end{figure*}


\begin{definition3}
\textbf{(Hypergraph Snapshots)}.
A snapshot of the hypergraph $\mathcal{G}=\{\mathcal{V},\mathcal{E}\}$ is a subgraph which can be defined as $\mathcal{G}_e=\{\mathcal{V}_e,\mathcal{E}_e\}$. Here $\mathcal{V}_e$ and $\mathcal{E}_e$ are the subsets of $\mathcal{V}$ and $\mathcal{E}$ respectively. Different from simple graph snapshots, hypergraph snapshots are generated according to hyperedge types, which means all the hyperedges in $\mathcal{E}_e$ should belong to the same hyperedge type. As shown in Figure \ref{fig:hyper_snap}, for example, there are three kinds of hyperedges in the original hypergraphs, and each hypergraph snapshot contains one type of hyeredges. 
\end{definition3}
\newtheorem{problem}{Problem}
\begin{problem}
\textbf{(Heterogenous Hypergraph Embedding for Graph Classification).}
Given a heterogeneous hypergraph $\mathcal{G}$, we aim to learn its representation $\mathbf{Z}^{\mathcal{G}} \in \mathbb{R}^{|\mathcal{V}|\times C}$, where each row of this matrix represents the embedding of each node. This representation can be used for downstream predictive applications such as nodes classification.
\end{problem}

\section{Heterogeneous Hypergraph Embedding}\label{sec:method}
%
The overview of our heterogeneous hypergraph embedding framework is shown in Figure \ref{fig:framework}. The input is a simple graph. If the simple graph is heterogenous, we first extract simple graph snapshots with different meta-paths.  Afterwards, we construct hypergraphs on these simple graphs and then decompose them into multiple hypergraph snapshots. We use our developed Hypergraph Wavelet Neural Network (\textbf{HWNN}) to learn node embeddings in each snapshot and then aggregates these snapshots into a comprehensive representation for the downstream classification.



\subsection{HWNN: Hypergraph Wavelet Neural Networks}\label{sub:HWNN}
\par For each vertex $v_i \in \mathcal{V}$, we first lookup its initial vector representation, $\mathbf{v}_i \in \mathbb{R}^{C\times 1}$, via a global embedding matrix and then project it into sub-spaces of different types of hyperedges. The representation of vertex $v_i$ in the hyperedge-specific space with hyperedge type $t_e \in \mathcal{T}_e$ is computed as:
\begin{equation}\label{equ:project}
	\mathbf{v}_i^{t_e}=\mathbf{M}_{t_e}\mathbf{v}_i
\end{equation}
where $\mathbf{M}_{t_e} \in \mathbb{R}^{C\times C}$ is the hyperedge-specific projection matrix of $t_e$. 

\subsubsection{Hypergraph convolution via Fourier basis}
For each snapshot $\mathcal{G}_e=\{\mathcal{V}_e,\mathcal{E}_e,\mathbf{W}\}$ extracted from the original heterogeneous hypergraph, the laplacian matrix is computed as $\mathbf{\Delta}^{\mathcal{G}_e}=\mathbf{I}-\mathbf{\Theta}^{\mathcal{G}_e}$, where $\mathbf{\Theta}^{\mathcal{G}_e}=(\mathbf{D}_v^{\mathcal{G}_e})^{-\frac{1}{2}}\mathbf{H}^{\mathcal{G}_e}\mathbf{W}(\mathbf{D}_e^{\mathcal{G}_e})^{-1}(\mathbf{H}^{\mathcal{G}_e})^T(\mathbf{D}_v^{\mathcal{G}_e})^{-\frac{1}{2}}$. Let $\mathbf{x}^{\mathcal{G}_e}_t(v_i)=\mathbf{v}_i^{t_e}(t)$. where $t$ is the index of elements in $\mathbf{v}_i^{t_e}$, $t=1,\cdots,C$, then $\mathbf{x}_t^{\mathcal{G}_e}=[\mathbf{v}_1^{\mathcal{G}_e}(t),\cdots,\mathbf{v}_{|\mathcal{V}|}^{\mathcal{G}_e}(t)]^T$. According to \cite{donnat2018learning}, the hypergraph Laplacian $\mathbf{\Delta}^{\mathcal{G}_e}$ is a $|\mathcal{V}|\times |\mathcal{V}|$ positive semi-definite matrix, it can be diagonalized as:
$$
\mathbf{\Delta}^{\mathcal{G}_e}=\mathbf{U}^{\mathcal{G}_e}\mathbf{\Lambda}^{\mathcal{G}_e}(\mathbf{U}^{\mathcal{G}_e})^T
$$
where $\mathbf{U}^{\mathcal{G}_e}$ is the Fourier basis, which contains the complete set of orthonormal eigenvectors ordered by its non-negative eigenvalues $\mathbf{\Lambda}^{\mathcal{G}_e}=\mathbf{diag}(\lambda_0,\cdots, \lambda_{n-1})$. According to the convolution theorem, the convolution operation $*_{hG}$ of $\mathbf{x}_t^{\mathcal{G}_e}$ and a filter $\mathbf{y}$ can be written as the Fourier inverse transform after the element-wise Hadamard product of their Fourier transforms:
\begin{equation}\label{equ:con_fourier}
\begin{split}
\mathbf{x}_t^{\mathcal{G}_e}*_{hG}\mathbf{y}&=\mathbf{U}^{\mathcal{G}_e}(((\mathbf{U}^{\mathcal{G}_e})^T\mathbf{x}^{\mathcal{G}_e})\odot ((\mathbf{U}^{\mathcal{G}_e})^T\mathbf{y}))\\
&=\mathbf{U}^{\mathcal{G}_e}\mathbf{\Lambda}_\theta^{\mathcal{G}_e} (\mathbf{U}^{\mathcal{G}_e})^T\mathbf{x}_t^{\mathcal{G}_e}
\end{split}
\end{equation}
Here $(\mathbf{U}^{\mathcal{G}_e})^T\mathbf{y}=[\theta_0,\cdots,\theta_{n-1}]^T$ is the Fourier transform of the filter, and $\mathbf{\Lambda}_\theta^{\mathcal{G}_e}=\mathbf{diag}(\theta_0,\cdots, \theta_{n-1})$. 

However, the above operation has the following two major issues. First, it is not localized in the vertex domain  \cite{xu2019graph}, which cannot fully empower the convolutional operation. Secondly, eigenvectors are explicitly used in convolutions, requiring the eigen-decomposition of the Laplacian matrix for each snapshot in $G$. To address these issues, we propose to replace the Fourier basis with a Wavelet basis. 

The rationale of choosing wavelet basis instead of original Fourier basis is as follows. First of all, the Wavelet basis is much sparser than the Fourier basis and most suitable for modern GPU architectures for efficient training \cite{neelima2012effective}. Moreover, by the nature of wavelet basis, the efficient polynomial approximation can be achieved more easily. Based on this feature, we thus able to further propose a polynomial approximate to the graph wavelets so that the eigen-decomposition of the Laplacian matrix is not needed anymore. Last but not least wavelets represent information diffusion process, which are very suitable for implementation of localized convolutions in the vertex domain, which has been theoretically proved and empirically validated in the recent studies \cite{xu2019graph, donnat2018learning}. Next, we introduce the details of altering this basis.

\subsubsection{Hypergraph convolution based on wavelets} With the above discussion, 
let $\mathbf{\psi}_s^{\mathcal{G}_e} = \mathbf{U}^{\mathcal{G}_e}\mathbf{\Lambda}^{\mathcal{G}_e}_s(\mathbf{U}^{\mathcal{G}_e})^T$ be a set of wavelets with scaling parameter $-s$. Here $\mathbf{\Lambda}^{\mathcal{G}_e}_s=\mathbf{diag}(e^{-\lambda_0s},\cdots,e^{-\lambda_{n-1}s})$ is the heat kernel matrix, and $\lambda_0 \leq \lambda_1 \leq \cdots \leq \lambda_{n-1}$ are eigenvalues of hypergraph laplacian $\Delta^{\mathcal{G}_e}$. Then, the hypergraph convolution based on the Wavelet basis can be obtained from Equation (\ref{equ:con_fourier}) after replacing the Fourier basis with $\mathbf{\psi}_s^{\mathcal{G}_e}$:
\begin{equation}\label{equ:con_wavelet}
\begin{split}
\mathbf{x}_t^{\mathcal{G}_e}*_{hG}\mathbf{y}&=\mathbf{\psi}_s^{\mathcal{G}_e}(((\mathbf{\psi}_s^{\mathcal{G}_e})^{-1}\mathbf{x}^{\mathcal{G}_e})\odot ((\mathbf{\psi}_s^{\mathcal{G}_e})^{-1}\mathbf{y}))\\
&=\mathbf{\psi}_s^{\mathcal{G}_e}\mathbf{\Lambda}_\beta^{\mathcal{G}_e}(\mathbf{\psi}_s^{\mathcal{G}_e})^{-1}\mathbf{x}^{\mathcal{G}_e}
\end{split}
\end{equation}
where $(\mathbf{\psi}_s^{\mathcal{G}_e})^{-1}\mathbf{y}$ is the spectral transform of the filter, and $\mathbf{\Lambda}_\beta^{\mathcal{G}_e}=\mathbf{diag}(\beta_0,\cdots,\beta_{n-1})$. In the following, we further introduce the Stone-Weierstrass theorem \cite{donnat2018learning} to approximate graph wavelets without requiring the eigen-decomposition of the Laplacian matrix, making our method much more efficient.

\subsubsection{Stone-Weierstrass theorem and polynomial approximation}
Note that Equation (\ref{equ:con_wavelet}) still needs the eigen-decomposition of the hypergraph Laplacian matrix. As the wavelet matrix is much sparser than Fourier basis, we can easily achieve an efficient polynomial approximation according to the Stone-Weierstrass theorem \cite{donnat2018learning}, which states that the heat kernel matrix $\mathbf{\Lambda}^{\mathcal{G}_e}_s$ restricted to $\left[0,\lambda_{n-1} \right]$ can be approximated by:
\begin{equation}\label{equ:Stone_Weier}
	\mathbf{\Lambda}^{\mathcal{G}_e}_s= \sum_{k=0}^K\alpha_k^{\mathcal{G}_e} (\mathbf{\Lambda}^{\mathcal{G}_e})^k+r(\mathbf{\Lambda}^{\mathcal{G}_e})
\end{equation}
where $K$ is the polynomial order. $\mathbf{\Lambda}^{\mathcal{G}_e}=\mathbf{diag}(\lambda_0,\cdots, \lambda_{n-1})$ contains the eigenvalues of hypergraph Laplacian $\Delta^{\mathcal{G}_e}$, and $r(\mathbf{\Lambda}^{\mathcal{G}_e})$ is the residual where each entry has an upper bound:
\begin{equation}
|r(\lambda)| \leq \frac{(\lambda s)^{K+1}}{(K+1) !}
\end{equation}
Then, the graph wavelet is polynomially approximated by:
\begin{equation}\label{equ:poly}
	\mathbf{\psi}_s^{\mathcal{G}_e}\approx \sum_{k=0}^K\alpha_k^{\mathcal{G}_e} (\mathbf{\Delta}^{\mathcal{G}_e})^k
\end{equation}
Since $\mathbf{\Delta}^{\mathcal{G}_e}$ can be seen as a first-order polynomial of $\mathbf{\Theta}^{\mathcal{G}_e}$, Equation (\ref{equ:poly}) can be rewritten as:
\begin{equation}\label{equ:poly_re}
	\mathbf{\psi}_s^{\mathcal{G}_e}\approx \mathbf{\Theta}_\Sigma^{\mathcal{G}_e}=\sum_{k=0}^K\theta_k (\mathbf{\Theta}^{\mathcal{G}_e})^k
\end{equation}
Obviously, we can replace $-s$ in $\mathbf{\Lambda}^{\mathcal{G}_e}_s$ with $s$ so that $(\mathbf{\psi}_s^{\mathcal{G}_e})^{-1}$ can be simultaneously obtained. However, Equation (\ref{equ:poly_re}) places the parameter $s$ into the residual item, which can be ignored if we take $s$ as a small value. Therefore, we use a set of parallel parameters to approximate $(\mathbf{\psi}_s^{\mathcal{G}_e})^{-1}$ as:
\begin{equation}\label{equ:poly_re_inv}
	(\mathbf{\psi}_s^{\mathcal{G}_e})^{-1}\approx (\mathbf{\Theta}_\Sigma^{\mathcal{G}_e})^{'}=\sum_{k=0}^{K^{'}}\theta_k^{'} (\mathbf{\Theta}^{\mathcal{G}_e})^k
\end{equation}
With the above transform, Equation (\ref{equ:con_wavelet}) can be deduced as:
\begin{equation}\label{equ:con_wavelet2}
\mathbf{x}_t^{\mathcal{G}_e}*_{hG}\mathbf{y}=\mathbf{\Theta}_\Sigma^{\mathcal{G}_e}\mathbf{\Lambda}_\beta^{\mathcal{G}_e}(\mathbf{\Theta}_\Sigma^{\mathcal{G}_e})^{'}\mathbf{x}^{\mathcal{G}_e}
\end{equation}
When we have a hypergraph signal $\mathbf{X}^{\mathcal{G}_e}=[\mathbf{x}_1^{\mathcal{G}_e},\cdots,\mathbf{x}_C^{\mathcal{G}_e}]$ with $|\mathcal{V}|$ nodes and $C$ dimensional features, our hyperedge convolution neural networks can be formulated as:
\begin{equation}\label{equ:con_final}
(\boldsymbol{X}_{[:, j]}^{\mathcal{G}_e})^{m+1}=h\left(\mathbf{\Theta_{\Sigma}}^{\mathcal{G}_e} \sum_{i=1}^{p} \boldsymbol{\Lambda}_{i, j}^{m} (\mathbf{\Theta_{\Sigma}}^{\mathcal{G}_e})^{'} (\boldsymbol{X}_{[:, i]}^{\mathcal{G}_e})^{m}\right)
\end{equation}
where $\boldsymbol{\Lambda}_{i, j}^{m}$ is a diagonal filter matrix, and $(\boldsymbol{X}^{\mathcal{G}_e})^{m} \in \mathbb{R}^{|\mathcal{V}|\times C_m}$ is the input of the m-th convolution layer. We can further reduce the number of filters by detaching the feature transform from the convolution, and Equation (\ref{equ:con_final}) can be simplified as:
\begin{equation}\label{equ:con_final_simple}
(\boldsymbol{X}^{\mathcal{G}_e})^{m+1}=h\left(\mathbf{\Theta_{\Sigma}}^{\mathcal{G}_e}  \boldsymbol{\Lambda}^{m} (\mathbf{\Theta_{\Sigma}}^{\mathcal{G}_e})^{'} (\boldsymbol{X}^{\mathcal{G}_e})^{m}\mathbf{W}\right)
\end{equation}
Where $\mathbf{W}$ is a feature project matrix. Let $\mathbf{Z}^{\mathcal{G}_e} \in \mathbb{R}^{|\mathcal{V}|\times C_{m+1}}$ be the output of the last layer $\mathbf{Z}^{\mathcal{G}_e}=(\boldsymbol{X}^{\mathcal{G}_e})^{m+1}$, then for all snapshots of $G=\{\mathcal{G}_1,\cdots,\mathcal{G}_{|\mathcal{T}_e|} \}$, we have graph representations as:
\begin{equation}\label{equ:z}
\mathbf{Z}=\mathbf{Z}^{\mathcal{G}_1}\oplus \cdots \oplus\mathbf{Z}^{\mathcal{G}_{|\mathcal{T}_e|}}
\end{equation}
Here $\oplus$ is the concatenation operation, and $\mathbf{Z}$ is the concatenation of $\mathbf{Z}^{\mathcal{G}_i},i=1,\cdots,|\mathcal{T}_e|$. Finally, the representation of the heterogeneous hypergraph $\mathcal{G}$ can be calculated by the summation over all its snapshots as:
\begin{equation}
\mathbf{Z}^{\mathcal{G}}=f(\mathbf{Z})
\end{equation}
where $f$ is a multilayer perceptron, and $\mathbf{Z}^{\mathcal{G}} \in \mathbb{R}^{|\mathcal{V}|\times C_{m+1}}$. In the task of node classification, $C_{m+1}$ should be equal to the number of classes. The loss function can be combined with the cross-entropy error over all labeled examples and the regularizer on projection matrices:
\begin{equation}
\mathcal{L}=-\sum_{v \in \mathcal{V}_{l}}\sum_{i=1}^{C_{m+1}}\mathbf{Y}_{v,i}\text{ln}\mathbf{Z}^{\mathcal{G}}_{v,i}+\eta \text{tr}(\mathbf{M}_{t_e}^T\mathbf{M}_{t_e})
\end{equation}
where $\mathcal{V}_{l}$ is the set of labeled nodes, $\mathbf{Y}_{v,i}$ is the label value of node $v$ in terms of the category $i$. If node $v$ belongs to category $i$, $\mathbf{Y}_{v,i}=1$, otherwise, 0. $\eta$ is a trade-off parameter of the regularizer. Here, we follow \cite{zhao2018personality} and use the trace of $\mathbf{M}_{t_e}^T\mathbf{M}_{t_e}$ as the regular term, which can be also replaced by the $L2$ regularization.

\subsection{Model Analysis and Discussion}\label{sub:connection}
\par In this section, we provide an analytical discussion about our model from multiple perspectives to show its advantages.

\subsubsection{$\mathbf{\Theta}^{\mathcal{G}_e}$ plays a role like an adjacency matrix}
\par Our method can achieve more profound learning results because we leverage the power of $\mathbf{\Theta}^{\mathcal{G}_e}$ for higher-order relations in hypergraphs, and $\mathbf{\Theta}^{\mathcal{G}_e}$ can be treated as an adjacency matrix of the hypergraph. 

As previously mentioned, we use $\mathbf{H} \in \mathbb{R}^{|\mathcal{V}|\times |\mathcal{E}|}$ to denote the presence of nodes in different hyperedges. If $v \in e$, we have $\mathbf{H}(v,e)=1$, and otherwise, the corresponding entries are set as $0$. In nature, $\mathbf{H}$ indicates the relations between nodes and hyperedges, then we can use $\mathbf{H}\mathbf{H}^T$ to describe connections between nodes. A normalized version can be written like: $\mathbf{H}\mathbf{W}\mathbf{D}_e^{-1}\mathbf{H}^T$. In order to remove self-loops in hypergraphs, we change the above formula by:
\begin{equation}\label{equ:adj_hyper}
	\mathbf{A}^h=\mathbf{H}\mathbf{W}\mathbf{D}_e^{-1}\mathbf{H}^T-\mathbf{D}_v
\end{equation} 
Then a normalized version of Equation (\ref{equ:adj_hyper}) can be rewritten by:
\begin{equation}
	\mathbf{A}^{normalized}=\mathbf{D}_v^{-\frac{1}{2}}\mathbf{H}\mathbf{W}\mathbf{D}_e^{-1}\mathbf{H}^T\mathbf{D}_v^{-\frac{1}{2}}-\mathbf{I}_v=\mathbf{\Theta}-\mathbf{I}
\end{equation} 
From the above formula, we can find $\mathbf{\Theta}$ has the similar meaning as the adjacency matrix.


\subsubsection{Higher-order relations for hypergraphs}
\par In order to elaborate on the advantages of our method, we first introduce a prior work \cite{chen2020hypergraph} on hypergraph neutral network, and then we discuss the relationships between our work and prior work. A simplified hypergraph convolution can be generated via extending simple graph convolution to the hypergraph. Recall that the typical GCN framework in simple graphs is defined as:

\begin{equation}\label{equ:gcn}
\mathbf{X}^{l+1}=\mathbf{D}^{-\frac{1}{2}}\mathbf{A}\mathbf{D}^{-\frac{1}{2}}\mathbf{X}^{l}\mathbf{W}
\end{equation}
Here $\mathbf{D}$ contains all nodes' degree of the graph. $\mathbf{A}$ is the adjacency matrix. Following Observation 2, we can effectively model hypergraph convolution in a similar way:
\begin{equation}\label{equ:hcn}
\mathbf{X}^{l+1}=\mathbf{\Theta}\mathbf{X}^{l}\mathbf{W}
\end{equation}
Here $\mathbf{X}^{l}$ is the signal at the $l$-th layer, and $\mathbf{W}$ is a feature projection matrix. The traditional convolutional neural network for simple graphs is a special case of this work because the Laplacian $\Delta$ can be degenerated as the simple graph Laplacian. 

When the filter $\mathbf{\Lambda}$ in Equation (\ref{equ:con_final_simple}) is initialized from value $1$, it is close to an  identity matrix $\mathbf{I}$. Let $K=1$ and $K^{'}=0$ for Equation (\ref{equ:poly_re}) and Equation (\ref{equ:poly_re_inv}) respectively, then Equation (\ref{equ:con_final_simple}) degenerates to Equation (\ref{equ:hcn}). That means we employ the polynomial of $\mathbf{\Theta}$ to extend prior works based on the hypergraph theory only, and this extension makes our method more profound for node representation learning. Since $\mathbf{\Theta}$ actually serves a similar role as an adjacency matrix, the power of $\mathbf{\Theta}$ can learn higher-order relations for hypergraphs. Furthermore, the filter $\mathbf{\Lambda}$ improves the performance in one more step via suppressing trivial components and magnifying rewarding components.

The complexity of Equation (\ref{equ:con_final_simple}) is $\mathcal{O}(N + pq)$ where $N$ is the number of nodes, $p$ and $q$ are input dimensions and output dimensions respectively. Inspired by the formula (\ref{equ:hcn}) and (\ref{equ:con_final_simple}), the complexity of our model can be further reduced to $\mathcal{O}(pq)$ with a simplified version:
\begin{equation}\label{equ:con_final_simple_plus}
(\boldsymbol{X}^{\mathcal{G}_e})^{m+1}=h\left(\sum_{k=0}^K (\mathbf{\Theta}^{\mathcal{G}_e})^k    (\boldsymbol{X}^{\mathcal{G}_e})^{m}\mathbf{W}\right)
\end{equation}
where $K$ is the polynomial approximation order, $\mathbf{W} \in \mathbb{R}^{p\times q}$ is feature transformation matrix.


\section{Experimental Results and Analysis}\label{sec:exp}
\par In this section, we introduce related setup for our experiment and discuss the evaluation restuls.

\subsection{Experimental Setting}
\subsubsection{Datasets}\label{subsec:data}
\begin{itemize}
	\item \textbf{Pubmed\footnote{ https://github.com/jcatw/scnn/tree/master/scnn/data/Pubmed-Diabetes
}}: The Pubmed dataset \cite{sen2008collective} contains 19, 717 academic publications with 500 features. These publications are treated as nodes and their citation relationships are treated as 44,338 links. Each node falls into one of three classes (three kinds of Diabetes Mellitus).  
	\item	\textbf{Cora\footnote{https://relational.fit.cvut.cz/dataset/CORA
}:} The Cora dataset \cite{mccallum2000automating} contains 2,708 published papers in the area of machine learning, which are divided into 7 categories (case based, genetic algorithms, neural networks, probabilistic methods, reinforcement learning, rule learning, and theory). There are 5,429 citation links in total. The paper nodes have 1,443 features. 
	\item \textbf{DBLP\footnote{http://web.cs.ucla.edu/~yzsun/Publications.html}:} The DBLP dataset \cite{sun2011pathsim} is an academic network from four research areas. There are 14,475 authors, 14,376 papers, and 20 conferences, among which 4,057 authors, 20 conference and 100 papers are labeled with one of the four research areas (database, data
mining, machine learning, and information retrieval). We use this dataset to predict the research areas of authors.
\end{itemize}
\par Note that the above four datasets cover all kinds of hypergraph heterogeneities we have mentioned: homogeneous simple graph with heterogeneous hypergraphs (Pubmed and Cora), and heterogeneous simple graph with heterogeneous hypergraphs (DBLP). 

\subsubsection{Baselines}
We choose six state-of-the-art graph and hypergraph embedding methods as baselines:
\begin{itemize}
	\item \textbf{Hypergraph Embedding Methods}
	\begin{itemize}
	\item \textbf{HyperEmbedding} \cite{zhou2007learning}. This method selects the top-$k$ eigenvectors derived from the Laplacian matrix of the hypergraph as the representation of the hypergraph.
	\item \textbf{HGNN} \cite{feng2019hypergraph}. It is a convolution network for hypergraphs based on the Fourier basis. The reason why we choose it as our baseline is that their reported performance exceeds a variety of hot methods such as GCN \cite{kipf2016semi}, Chebyshev-GCN \cite{defferrard2016convolutional}, and Deepwalk \cite{perozzi2014deepwalk}. 
	\end{itemize}
	\item \textbf{Simple Embedding Methods}
	\begin{itemize}
	\item \textbf{PME} \cite{chen2018pme}. It is a heterogenous graph embedding model based on the metric learning to capture both first-order and second-order proximities in a unified way. It learns embeddings by firstly projecting vertices from object space to corresponding relation space and then calculates the proximity between projected vertices.
	\item \textbf{HHNE} \cite{wang2019hyperbolic}. This method learns node embeddings in Hyperbolic space instead of Euclidean space. It follows the idea of word2vec \cite{mikolov2013distributed} and generates the corpus via bias random walk on the networks.  
	\item \textbf{metapath2vec} \cite{metapath2vec}. It extends the Deepwalk model to learn representations on heterogeneous networks through meta path guided random walks.  
	\item  \textbf{GWNN} \cite{xu2019graph}. It is a graph convolution network based on the Wavelet basis, but it is designed for simple homogeneous graphs. It was reported to beat GCN, Spectral CNN \cite{bruna2013spectral}, MoNet \cite{monti2017geometric}, and achieved the best results on homogeneous simple graphs. 
	\end{itemize}
\end{itemize}

\begin{figure*}[h]
\centering
\subfloat[Accuracy for Cora]{
\label{agree}
\includegraphics[width=0.25\textwidth]{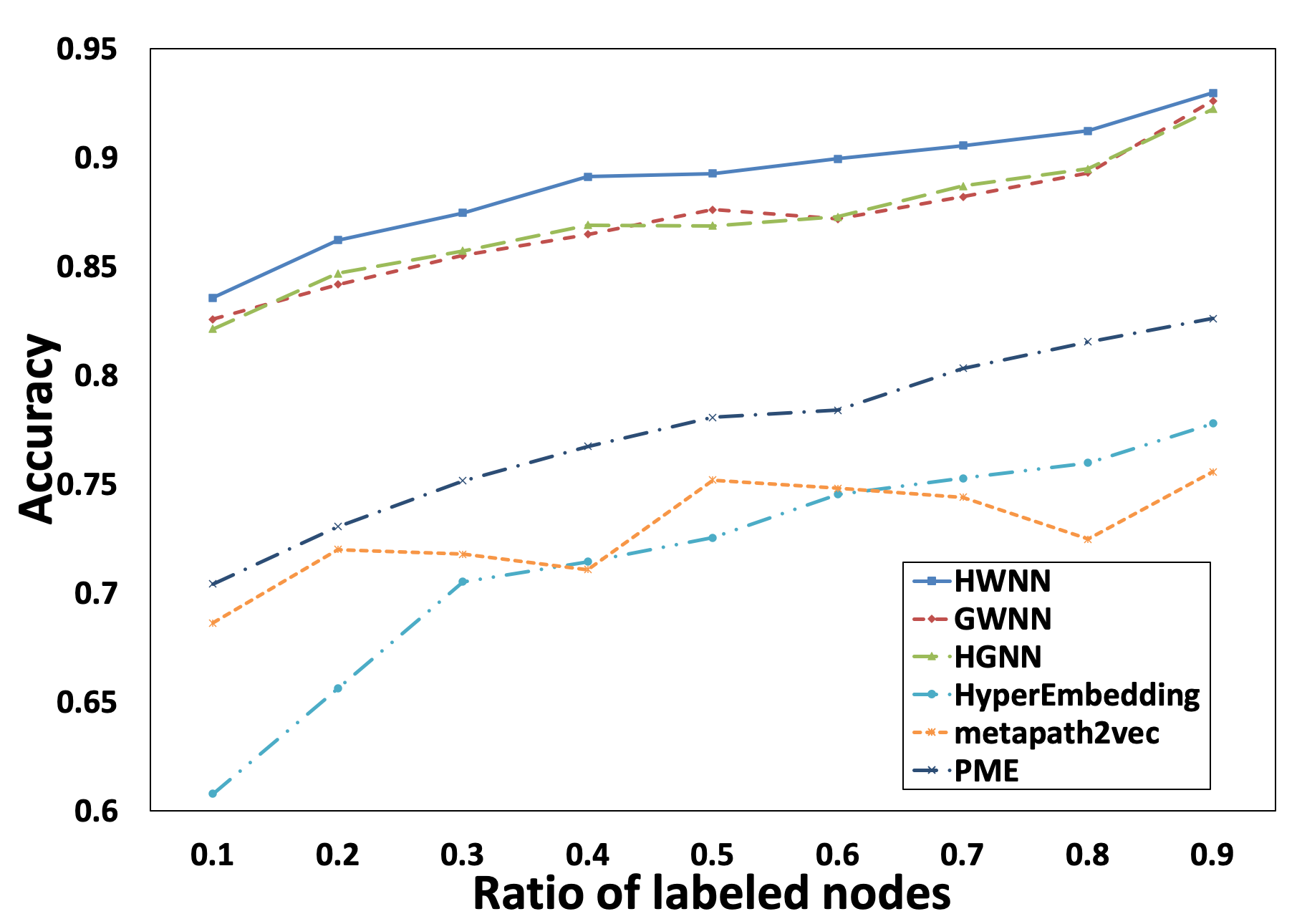}%
}
\subfloat[F1 for Cora]{
\label{ext}
\includegraphics[width=0.25\textwidth]{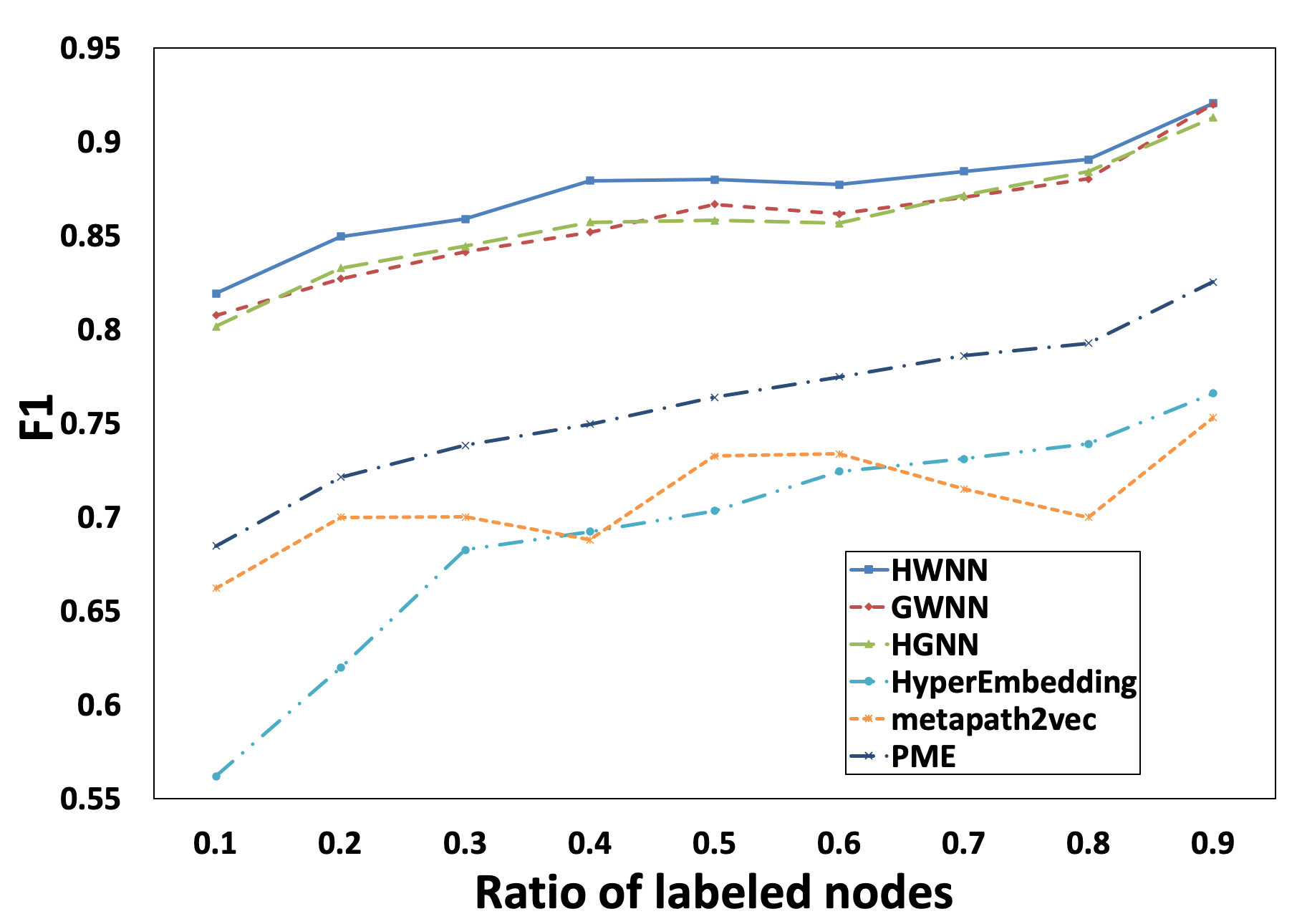}%
}
\subfloat[Precision for Cora]{
\label{ext}
\includegraphics[width=0.25\textwidth]{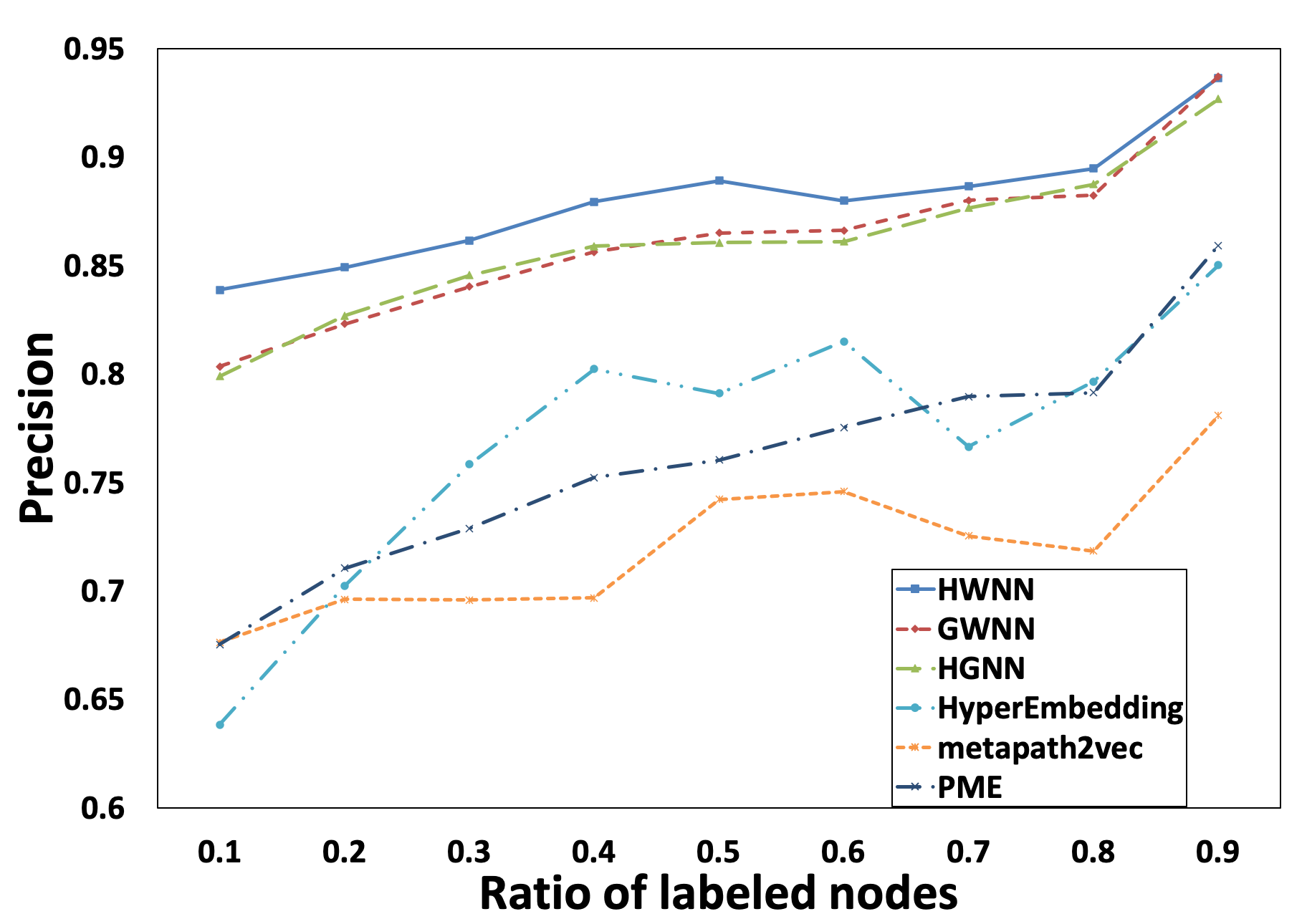}%
}
\subfloat[Recall for Cora]{
\label{ext}
\includegraphics[width=0.25\textwidth]{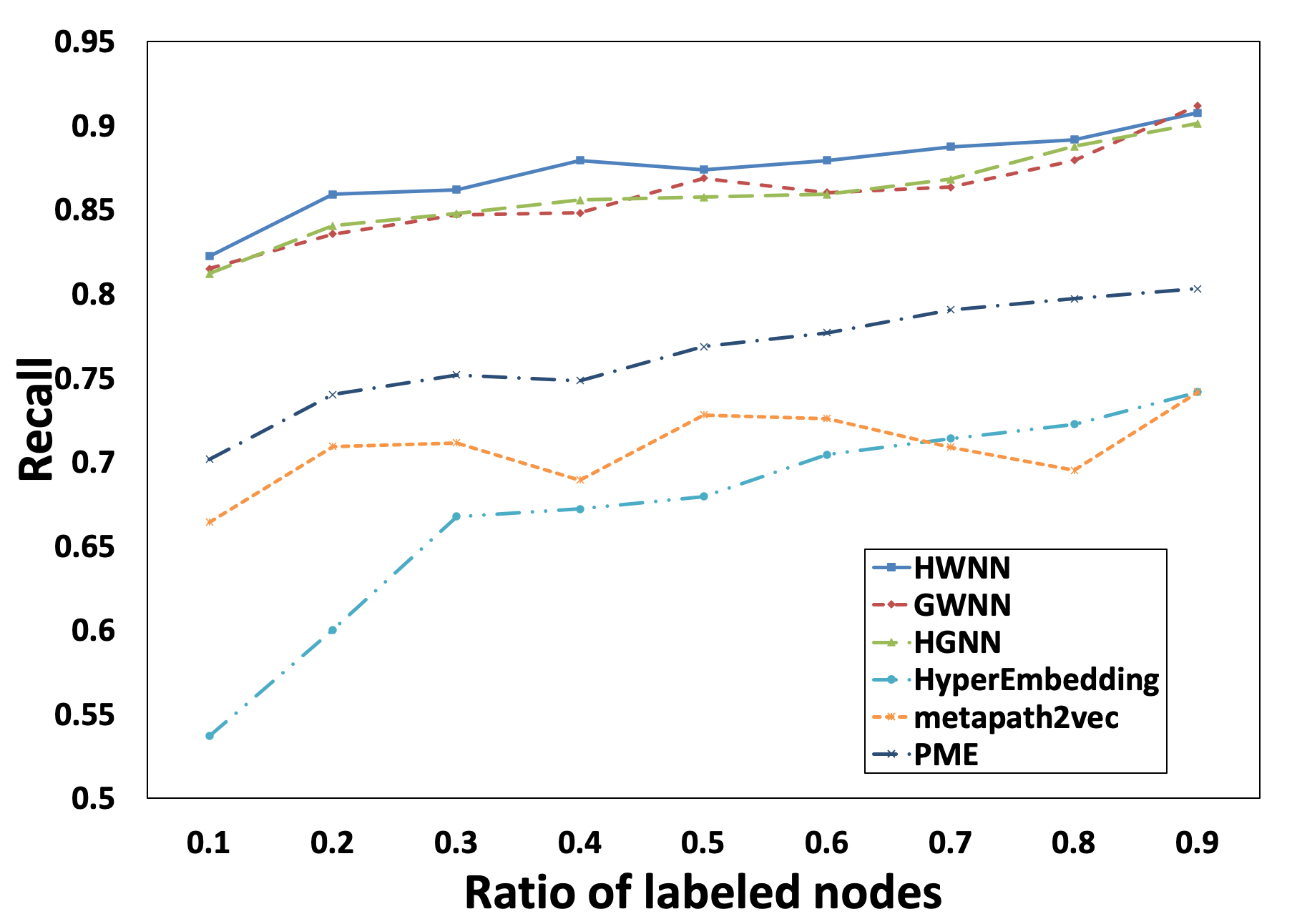}%
}
\caption{Results w.r.t. Ratio of labeled nodes.}
\label{fig:multi_label_classification}
\end{figure*}

\begin{table}[]
\centering
\caption{Results on node classification (50\% labeled)}
\label{tab:classification_results}
\resizebox{0.45\textwidth}{!}{%
\begin{tabular}{p{0.12\textwidth}p{0.05\textwidth}p{0.03\textwidth}p{0.045\textwidth}|p{0.045\textwidth}p{0.03\textwidth}p{0.04\textwidth}}
\toprule
 & \multicolumn{3}{c}{Recall (\%) }& \multicolumn{3}{|c}{F1 (\%) }\\ \midrule
Methods  & Pubmed &  Cora & DBLP  & Pubmed &  Cora & DBLP\\ \midrule
\textbf{HWNN}&\textbf{85.59}&\textbf{88.52}&\textbf{90.03}&\textbf{85.15}&\textbf{88.60}&  \textbf{91.45}\\
GWNN & 81.02  &   85.12 &  87.23 &  80.93 & 85.04  &  87.20 \\
HyperEmbedding&59.27& 54.01&54.22&60.31&57.84&54.59\\
HGNN &  82.32 &   82.87 & 86.34  &  82.08  &    84.09  &  85.47 \\
PME & 55.17  &    76.49  & 82.75  &  55.06  &    75.94 &  78.42 \\
HHNE   &  76.86   & 70.74  &  86.66 &  79.94   &  69.95 &  86.03 \\
metapath2vec &  74.48  &  66.68 & 82.22  &  74.97&   68.13 &  82.20 \\
random guess & 34.67   & 14.47  & 34.79  & 30.48   & 10.33  &32.75  \\

\midrule
& \multicolumn{3}{c}{Accuracy (\%) }& \multicolumn{3}{|c}{Precision (\%) }\\ \midrule
Methods  & Pubmed &  Cora & DBLP  & Pubmed &  Cora & DBLP\\ \midrule
\textbf{HWNN}   & \textbf{86.48}    & \textbf{89.73}  & \textbf{93.41}  &   \textbf{84.78} &    \textbf{89.12} &  \textbf{93.61} \\
GWNN& 82.51  &  85.82 &  90.11 &   80.88   &     85.20 &  90.50 \\
HyperEmbedding&67.49 &   64.55 &  74.73 &  78.32   &  79.35 &  87.63 \\
HGNN & 83.51 & 85.82 & 89.01  & 82.19 &    85.95  &  86.13 \\
PME  & 59.21  &      77.25& 80.22  &  55.40  &   75.86 &  76.07 \\
HHNE  & 81.32  &  71.64 & 83.24  &   81.32 &   75.95 & 87.32  \\
metapath2vec&76.86 &   70.31 &  86.81 &  75.96 &  70.72 &  87.36 \\
random guess & 42.45  & 31.09  & 48.35  & 37.55   & 11.58  & 41.59  \\
 \bottomrule
\end{tabular}
}
\end{table}

\subsubsection{Parameter Settings}
\par The optimal hyper-parameters in all the models are selected by the grid search method on the validation set. Our model achieves its best performance with the following setup. There are two convolution layers. The activation function of the first layer is ReLU, and the second layer is Softmax. The feature dimension of the hidden layer is set as 64, with a drop rate of 0.5. We use Adam optimizer with learning rate 0.001, regularization coefficient 0.0001, and 400 epochs. For PME, metapath2vec, and HHNE, we choose three meta-paths on DBLP dataset: P-C (paper-conference), P-A (paper-author), and A-P-A. (paper-conference-author). Other datasets have homogeneous simple graphs so we just use V-V (vertex-vertex) as the meta-path. For PME, we set $m=100$, and the number of negative samples for each positive edge is set $K=10$. The initial nodes features are generated from nodes' attributes provided by these datasets. 
\subsubsection{Hypergraph Construction}\label{subsec:hyper_con}
\par We define the following four types of hyperedges to construct the corresponding heterogeneous hypergraphs for the datasets:
\begin{itemize}
	\item \textbf{Neighbor-based hyperedges.} We first obtain each node's $\phi$-hop neighbors and connect them in one hyperedge. $\phi$ is set as 3 in our experiments. All these hyperedges belong to the same hyperedge type if the simple graph is homogeneous. If the simple graph is heterogeneous, a node has different types of neighbors following different meta-paths, and thus we can generate different types of hyperedges.
	\item \textbf{Attribute-based hyperedges.} Let $\mathcal{A}$ denote all the discrete attributes. For each attribute $a \in \mathcal{A}$, it has $ c_a$ values. Then, we build a type of hyperedges for each attribute $a$, and each hyperedge connects the nodes sharing the same attribute value. So there are $c_a$ hyperedges generated for attribute $a$. Each hyperedge type is treated as a hypergraph snapshot, and we finally generate $|\mathcal{A}|$ hypergraph snapshots accordingly.
	\item \textbf{Cluster-based hyperedges.} Given each node's feature vectors, we first calculate the cosine similarity between each pair of nodes and obtain several clusters by K-means. Then each cluster serves as a hyperedge. All the hyperedges belong to the same hyperedge type.
	\item \textbf{Community-based hyperedges.} We use the Louvain algorithm \cite{blondel2008fast} to discover communities in the corresponding simple graph via modularity maximization. Then each community is treated as a hyperedge. All these hyperedges belong to the same hyperedge type.
\end{itemize}

\subsection{Overall Performance in Node Classification}\label{subsec:classification}
 The comparison results in terms of F1, Recall, Accuracy, and Precision are reported in Table \ref{tab:classification_results}, which show that our solution consistently outperforms all baselines on all the datasets. Note that the differences between our solution and the other comparison methods are statistically significant ($p < 0.01$). Specifically, on the Pubmed dataset, our method exceeds the baselines at least 4.0\% in Recall, 3.7\% in F1, 3.6\% in Accuracy, and 3.2\% in Precision. On the Cora dataset, our solution also outperforms all baselines, and the relative improvements are 4.0\% (Recall), 4.2\% (F1), 4.6\% (Accuracy), and 3.7\% (Precision). On the DBLP dataset, the relative improvements achieved by our model are at least 3.2\% (Recall), 4.9\% (F1), 3.7\% (Accuracy), and 3.4\% (Precision). 
 

\begin{table}[t]
\centering
 \caption{Performance on Cora w.r.t $K$ (30\% labeled)}
 \label{tab:K}
 \resizebox{0.35\textwidth}{!}{%
 \begin{tabular}{p{0.03\textwidth}p{0.07\textwidth}p{0.05\textwidth}p{0.09\textwidth}p{0.09\textwidth}}
  \toprule
   & Recall (\%) & F1 (\%) & Accuracy (\%)& Precision (\%) \\
  \midrule
 K=1 &70.77 & 72.53 & 74.30 & 84.46\\
   \midrule
 K=2 & 86.75 & 87.53 & 89.59 & 88.71\\
    \midrule
 K=3 & 86.17 & 87.84 & 89.07 & 88.68\\
   \midrule
K=$\infty$ & 86.44 & 87.10 & 89.37 & 88.54\\
  \bottomrule
 \end{tabular}
 }
 \begin{tablenotes}
        \footnotesize
        \item[] We use K=$\infty$ to denote the case without approximation.
 \end{tablenotes}
\end{table}


\par \textbf{Impact of Label Sparsity}. We further study the impact of the label sparsity (i.e., label ratio) by varying the ratio of labeled nodes from 10\% to 90\%. We only report the results on the Cora dataset and similar comparison results/trends are also observed on the other datasets. Figure \ref{fig:multi_label_classification} shows the comparison results on the Cora dataset, from which we can observe that even in a lower label ratio, our method still outperforms the baselines, which is particularly useful because most real world datasets are sparsely labeled.

\subsection{Analysis of the Polynomial Approximation to the Wavelet Basis}\label{subsec:sens}
\par In this section, we study the impact of the polynomial approximation order $K$ on the classification performance. Since the Wavelet basis is much sparser than the Fourier basis, we develop an efficient polynomial approximation method to approximate it. In this experiment, we set different orders of the polynomial approximation to the Wavelet basis and report their performance in Table \ref{tab:K}. When $K$ increases from 1 to 2, the performance of our model climb significantly. However, when $K$ increases from 2 to 3, the performance has become steady and is very close to the original Wavelet basis without any approximation, which indicates that our model training can be super-efficiently done without any performance compromise. Please note that when $K=2$, both $\mathbf{\Theta_{\Sigma}}^{\mathcal{G}_e}$ and $(\mathbf{\Theta_{\Sigma}}^{\mathcal{G}_e})^{'}$ in Equation (\ref{equ:con_final_simple}) are set as 2 orders.

\begin{figure}[h]
\centering
\includegraphics[width=0.28\textwidth]{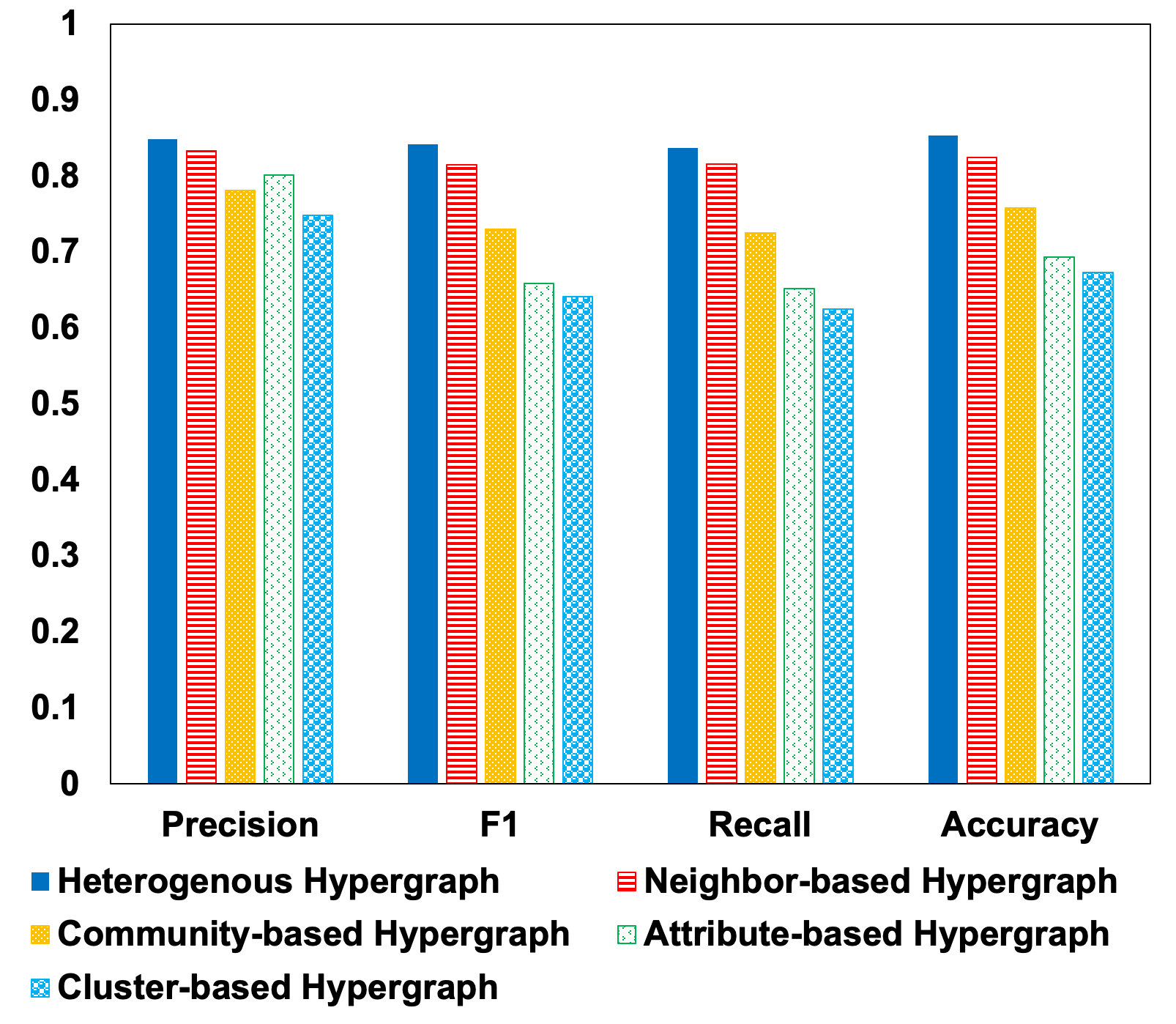}
\caption{Hypergraph snapshots on Cora (10\% labeled).}
\label{fig:hyper_const}
\end{figure}


 
%

\subsection{Impact of Hypergraph Construction}
\par As we propose four general hypergraph construction methods which do not depend on domain knowledge, we study the performance of each type of constructed hypergraphs and their jointly formed heterogeneous hypergraph in this experiment. Figure \ref{fig:hyper_const} shows the nodes classification results on Cora dataset, and we observe that the best performance is achieved on the heterogeneous hypergraph, showing the advantage of exploiting heterogeneous hypergraphs over the homogeneous hypergraphs. Another that the community-based hypergraph has only 28 hyperedges but achieve comparable performance with neighbor-based snapshot (1000 hyperedes). This indicates the potential superiority of the hypergraph-based method when dealing with large-scale networks because the number of hyperedges is much smaller than the numbers of nodes and edges on the simple graphs.

\subsection{Training Efficiency}
\par
\begin{figure}[h]
\centering
\includegraphics[width=0.28\textwidth]{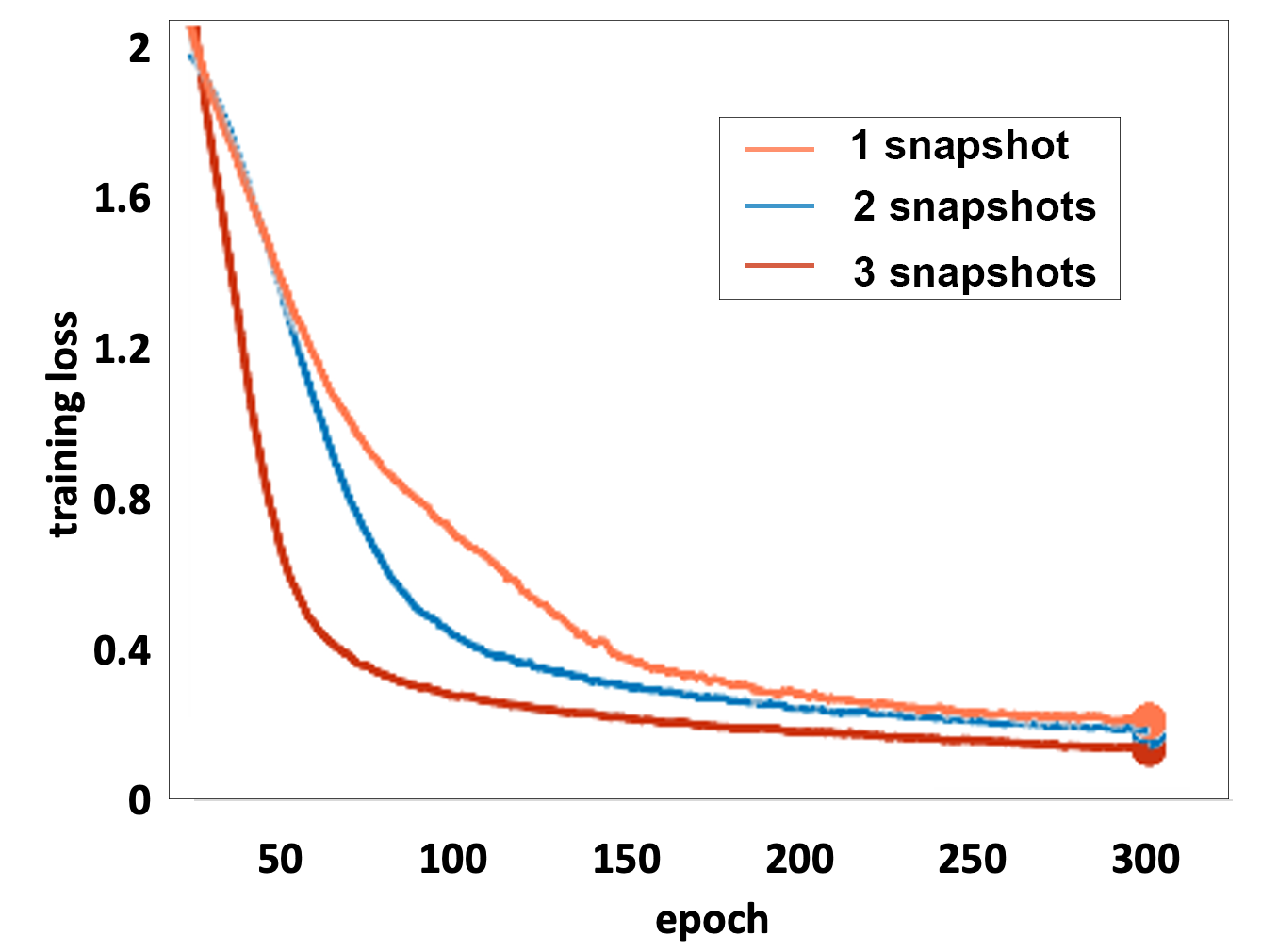}
\caption{Training loss w.r.t snapshot numbers.}
\label{fig:train_performance_size}
\end{figure}

 \textbf{Impact of snapshot number}. In order to evaluate the training efficiency of our model, we first study the convergence rate of our model training with different number of hypergraph snapshots. As shown in Figure \ref{fig:train_performance_size}, our model converges faster and the training loss becomes lower when it processes more snapshots. That means more available snapshots are helpful for our model to fast learn high-quality node embedding, thus accelerating the model convergence.

\noindent \textbf{Efficiency of the polynomial approximations.} In order to evaluate computation time of our model and Fourier-based model on the experiment datasets. We record the model training time in 100 epoch and report the results in Table \ref{tab:com_time}. Our model shows efficient training performance. Here we let the order of polynomial approximation equal to 4, which means both $\mathbf{\Theta_{\Sigma}}^{\mathcal{G}_e}$ and $(\mathbf{\Theta_{\Sigma}}^{\mathcal{G}_e})^{'}$ in Equation (\ref{equ:con_final_simple}) are set as 4 orders. The simplified version has much faster speed, and considering hypergraph snapshots can be assembled flexibly, we can leverage distributed computation for even larger datasets.

\begin{table}[h]
\centering
 \caption{Model Training Time (second)}
 \label{tab:com_time}
  \resizebox{0.3\textwidth}{!}{%
 \begin{tabular}{p{0.15\textwidth}p{0.05\textwidth}p{0.06\textwidth}p{0.05\textwidth}}
  \toprule
  & Cora & Pumbed & DBLP \\
   \midrule
Our method & 0.91  &433.14 &526.83\\
    \midrule
Fourier basis method &1.41& 632.76	&764.73	\\
   \midrule
speed-up ratio & 1.55	&1.46	&1.45	\\
  \bottomrule
 \end{tabular}
 }
\end{table}

\subsection{Case Study: Online Spammer Detection}\label{subsection:spammer}
\par Online spammer detection is an important research area. Previous researches \cite{liu2016pay, wu2015social, sunspammer2020} usually take pairwise relationships into considerations, ignoring non-pairwise relationships, and thus their performance can be further improved by applying our method here. 
In this section, we apply our method into spammer detection on the Spammer dataset from Liu et al. \cite{sunspammer2020}. The dataset consists of 14, 774 users in Sina platform\footnote{\url{https://www.sina.com.cn}}, and 3, 883 users are labeled as spammer or normal users. Spammers take up to 23.3\% of the total labeled users. 
We construct an adjacency matrix according to the following edges and each row serves as the first part of the feature vector. The remaining features come from the feature collection calculated by  \cite{sunspammer2020}, including folllowee/follower number, lifetime, mutual friends, number of tweets and so on. All these features are concatenated and then fed into an autoencoder to generate a feature matrix with 100 dimensions. We construct the hypergraph snapshots according to the second part of features, the similarity of all features, and the topological neighbors.  
\begin{figure}[h]
\centering
\includegraphics[width=0.25\textwidth]{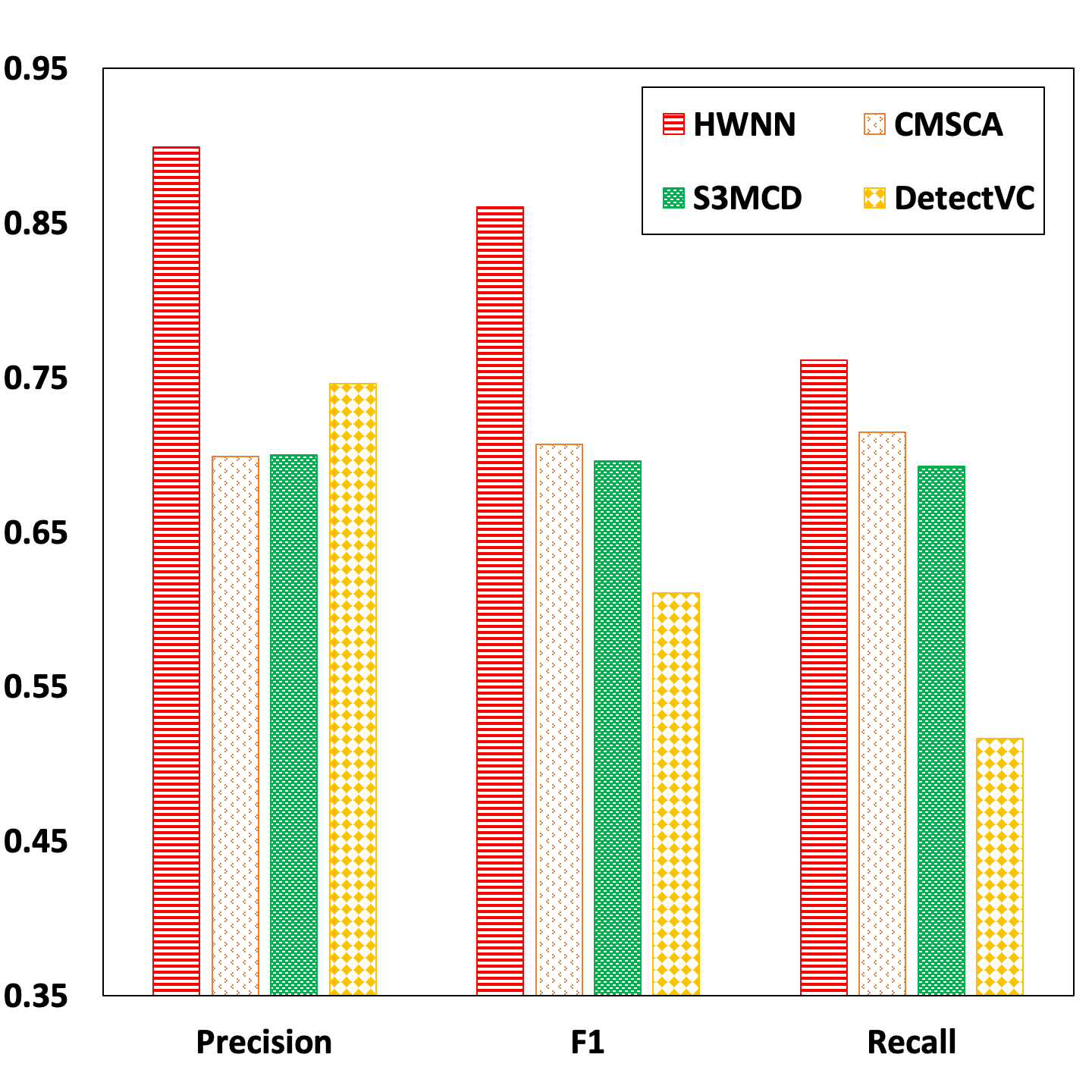}
\caption{Performance of Spammer Detection.}
\label{fig:spammer}
\end{figure}
\par We compared our method with other famous baselines like S3MCD \cite{liu2016pay}, DetectVC \cite{wu2015social}, and CMSCA \cite{sunspammer2020}. Results are reported in Figure \ref{fig:spammer}, from which we can demonstrate the validity of our method for the spammer detection. Compared with other baselines, HWNN achieved an average of 86.02\% in the F1 score, 89.88\% in the Precision score, and 76.11\% in the Recall score, exceeding to the second by 21.8\%, 20.5\%, and 6.6\% respectively. These results point out a new direction when studying spammer detection, that considering relationships beyond pairwise can further improve the final performance.

\section{Related work}\label{sec:relatedwork}
In this section, we briefly review simple graph embedding methods and hypergraph embedding methods.

\textbf{Simple Graph Embedding Methods.}
 Earlier methods for simple graph embedding such as LINE \cite{line}, DeepWalk \cite{perozzi2014deepwalk}, Node2Vec \cite{node2vec}, Metapath2vec \cite{metapath2vec}, PTE \cite{tang2015pte} mainly focus on capturing neighbourhood proximities among vertices. Recently, graph convolutional neural networks have achieved remarkable performance because of their powerful representation ability. Existing methods can fall into two categories, spatial methods and spectral methods. Spatial methods build their frameworks as information propagation and aggregation. Recently, various neighborhood selection methods and aggregation strategies have been proposed. In the work of Niepert et al. \cite{niepert2016learning}, the number of neighborhood nodes is fixed according to pairwise proximity. Hamilton et al. \cite{hamilton2017inductive} put forward a GraphSAGE method, where neighboring nodes are selected through sampling and then fed into an aggregate function. Petar et al. \cite{velivckovic2017graph} learn the aggregate function via the self-attention mechanism. Spectral methods formulate graph convolution in the spectral domain via graph Fourier transform \cite{bruna2013spectral}. To make the method computationally efficient and localized in the vertex domain, Defferrard et al. \cite{defferrard2016convolutional} parameterize the convolution filter via the Chebyshev expansion of the graph Laplacian. Xu et al. \cite{xu2019graph} further take Wavelet basis instead of the Fourier basis to make the graph convolution more powerful. 
However, existing graph neural networks only consider pairwise relationships between objects, and they cannot apply to non-pairwise relation learning. Recently, hypergraph methods have been emerging.

\textbf{Hypergraph Embedding Methods.}
%
 Some researchers construct a hypergraph to improve the performance of various tasks like video object segmentation \cite{huang2009video}, cancer prediction \cite{hwang2008learning}, and multi-modal fusion \cite{gao2012visual, zhao2018personality}. Most of them add a hypergraph regularizer into their final objective functions, and the regularizer is usually the same as in the spectral hypergraph partitioning  \cite{zhou2007learning}, in which the representations of connected nodes should be closer than those without links \cite{zheng2018novel}. Unlike the above works, Chu et al. \cite{chu2018social} learn their hypergraph embeddings through the analogy between hyperedges and text sentences. Then they maximize the log-probability of the centroid generated by its context vertices. In recent years, some works have  explored the hypergraph learning via hypergraph neural networks. Feng et al. \cite{feng2019hypergraph} introduced the concept of hypergraph Laplacian, and then proposed a hypergraph version of graph convolution based on simple graph convolution  \cite{defferrard2016convolutional}. Jiang et al. \cite{jiang2019dynamic} propose a hyperedge convolution method so that they learn a dynamic hypergraph. Yadati et al. \cite{yadati2019hypergcn} treat a hyperedge as a new kind of node and then the hypergraph can be changed as a simple graph. Tu et al. \cite{tu2018structural} design a framework for hypergraphs to preserve the first-order proximity and the second-order proximity of hypergraphs.

\section{Conclusion}\label{sec:con}
\par In this paper, we proposed a novel graph neural network for heterogeneous hypergraphs with a Wavelet basis. We further propose a polynomial approximation-based wavelets so that we can avoid the time-consuming Laplacian decomposition. Extensive evaluations have been conduct and experimental results demonstrate the advantage of our method. 

\section{ACKNOWLEDGMENTS}
This work is supported by National Key R\&D Program of China (Grant No. 2017YFB1003000, 2019YFC1521403), National Natural Science Foundation of China (Grants No. 61972087, 61772133,  61632008, 61702015, U1936104), National Social Science Foundation of China (Grants No. 19@ZH014),  Natural Science Foundation of Jiangsu province (Grants No. SBK2019022870), Jiangsu Provincial Key Laboratory of Network and Information Security (Grant No. BM2003201), Key Laboratory of Computer Network Technology of Jiangsu Province (Grant No. BE2018706), Key Laboratory of Computer Network and Information Integration of Ministry of Education of China (Grant No. 93K-9), China Scholarship Council (CSC), the Fundamental Research Funds for the Central Universities 2020RC25, Australian Research Council (Grant No. DP190101985, DP170103954). The first author Mr. Xiangguo Sun, in particular, wants to thank his parents for their support during his tough period.

\bibliographystyle{ACM-Reference-Format}
\bibliography{cssRef}

\end{document}